\documentclass{article}

\usepackage[preprint, nonatbib]{neurips_2024}

\usepackage{graphicx}

\usepackage[utf8]{inputenc} % allow utf-8 input
\usepackage[T1]{fontenc}    % use 8-bit T1 fonts
\usepackage{hyperref}       % hyperlinks
\usepackage{url}            % simple URL typesetting
\usepackage{booktabs}       % professional-quality tables
\usepackage{amsfonts}       % blackboard math symbols
\usepackage{nicefrac}       % compact symbols for 1/2, etc.
\usepackage{microtype}      % microtypography
\usepackage{xcolor}         % colors
\usepackage{listings}

\title{Real-time Position Reconstruction for the KamLAND-Zen Experiment using Hardware-AI Co-design}

\author{
  Alexander Migala \\
  Department of Physics\\
  University of California, San Diego\\
  La Jolla, CA 92037 \\
  \texttt{amigala@ucsd.edu} \\
  \And
  Eugene Ku \\
  Department of Physics \\
  University of California, San Diego\\
  La Jolla, CA 92037 \\
  \And
  Zepeng Li \\
  Department of Physics \\
  University of California, San Diego\\
  La Jolla, CA 92037 \\
  \And
  Aobo Li \\
  Department of Physics \\
  University of California, San Diego\\
  La Jolla, CA 92037 \\
  \texttt{aol002@ucsd.edu} \\
}

\lstnewenvironment{myverbatim}[1][]{
    \lstset {
        basicstyle=\ttfamily,
        frame=tb,
        #1,
        breaklines=True,
        columns=flexible
    }
}{}
\usepackage{amsmath}
\newcommand{\rfsoc}{RFSoC~4$\times$2~}
\DeclareMathOperator{\mse}{MSE}

\begin{document}
\bibliographystyle{abbrv}

\maketitle

\begin{abstract}
% If neutrinoless double beta decay, which is a rare decay phenomenon, were to be experimentally observed, would change the way we view the Standard Model.

% While there are several experiments searching for neutrinoless double beta decay $(0\nu\beta\beta)$, a rare decay phenomenon, we focus our attention on the KamLAND-Zen experiment, which is a monolithic liquid xenon scintillator detector. Because the events that occur in the detector are indirectly observed, a reconstruction of the event must be done, which is a non-trivial process and is done after all the data is collected. This means that the experiment commits resources towards storing data that is not relevant to the $0\nu\beta\beta$ analysis. To help solve this problem, we propose a toolchain that deploys a machine learning model, PointNet, onto a Field Programmable Gate Array so that a fast reconstructence may be done at the time of data collection. This is the first time that hardware-Algorithm co-design has been brought to the stage of $0\nu\beta\beta$ experiments.
Monolithic liquid scintillator detector technology is the workhorse for detecting neutrinos and exploring new physics. The KamLAND-Zen experiment exemplifies this detector technology and has yielded top results in the quest for neutrinoless double-beta ($0\nu\beta\beta$) decay. To understand the physical events that occur in the detector, experimenters must reconstruct each event's position and energy from the raw data produced. Traditionally, this information has been obtained through a time-consuming offline process, meaning that event position and energy would only be available days after data collection. This work introduces a new pipeline to acquire this information quickly by implementing a machine learning model, PointNet, onto a Field Programmable Gate Array~(FPGA). This work outlines a successful demonstration of the entire pipeline, showing that event position and energy information can be reliably and quickly obtained as physics events occur in the detector. This marks one of the first instances of applying hardware-AI co-design in the context of $0\nu\beta\beta$ decay experiments.
\end{abstract}

\section{Introduction \label{sec:intro}}
% In the field of particle physics, the Standard Model is the current working theory that best describes the foundational building blocks of our universe. Among the list of these foundational building blocks are the neutrinos. However, the Standard Model offers an incomplete view of neutrinos: namely that it is unknown if they are their own antiparticle. If they are, they would be designated as Majorana particles. If this behavior were ever observed, this would have enormous consequences on the Standard Model and neutrino mass. One way to experimentally probe whether neutrinos are Majorana particles is to search for neutrinoless double beta decay ($0\nu\beta\beta$), which is a process that involves a radioactive sample that decays in a specific way into an ion and two electrons. Experimentalists in the field have already begun efforts to experimentally search for such a process.
Liquid scintillator technology \cite{borex_agostini_experimental_2020} \cite{albanese_sno_2021} \cite{klz_search_for_majorana_nature} \cite{kamland_oscillation_result} \cite{juno_result} \cite{borexino_directionality_result} has been at the heart of recent searches for neutrino physics including solar neutrinos and neutrinoless double beta decay, a rare decay phenomenon. This technology probes for physics by emitting photons through energy excitations of the scintillator. The KamLAND-Zen (KLZ) experiment has used this technology to produce the world-leading results in the search for $0\nu\beta\beta$ decay \cite{klz_search_for_majorana_nature}. KLZ features a spherical tank of liquid xenon scintillator, which is surrounded by a spherical array of 1,879 inner detector photomultiplier tubes (PMTs) and 247 outer detector PMTs (used for background rejection), as shown in Figure~\ref{fig:flowchart} LEFT. 
%Because liquid scintillation technology does not allow for a direct representation of the neutrino physics inside the detector, experimenters must reconstruct the physics from the map of PMT data, which can often take days to fully analyze. 
%This data has five labels: the two PMT readout values and the three spatial coordinates of the PMT's location in the detector. 
When a physics event occurs in the detector, it emits isotropic scintillation light. Each PMT capturing the scintillation light will provide two key readouts: the arrival time of the light and the integrated charge. Consequently, a single data point in KamLAND-Zen is represented as a point cloud with 2,126 points, where each point consists of five dimensions: the two PMT readout values plus the x, y, and z coordinates of the PMT's location. To search for neutrinoless double-beta decay, researchers must extract essential physical information, such as the event position within the detector and the event energy. This process is known as event reconstruction.

Traditionally, event reconstruction is performed offline. This means that data is first collected without any reconstruction over a period of time (usually $\mathcal{O}(1~day)$) and stored in an offline storage facility. A fitting algorithm is then applied to the stored data to generate the event's position and energy. Due to the episodic nature of offline reconstruction, neither event position nor energy information is available until $\mathcal{O}(1~day)$ later. In this paper, a pipeline for deploying a machine learning algorithm (PointNet) onto a Field Programmable Gate Array (FPGA)--a specialized type of computer used for reducing processing overhead--for online fast reconstruction of KLZ detector events is proposed. Since KLZ will be equipped with 120 data acquisition boards with \rfsoc{FPGA} chips, deploying algorithms onto FPGAs could enable real-time retrieval of event position and energy information. 
%This pipeline begins at the high-level machine learning algorithm to the PMT data and ends at the deployment of the algorithm onto an FPGA. 
This pipeline relies on a key software package, \texttt{cgra4ml} \cite{abarajithan2024cgra4mlframeworkimplementmodern}, which allows the conversion from a machine learning design into its hardware representation. This work demonstrates the capabilities of this pipeline by deploying PointNet onto a single \rfsoc{FPGA}. Section \ref{sec:methods} will discuss the proposed pipeline for deploying the algorithm to hardware and the decisions made regarding the application of machine learning. Section \ref{sec:results} will demonstrate the reconstruction results from the PointNet model and the latency of the deployed model. Lastly, Section \ref{sec:summary} will summarize this toolchain and the applications of such a system.

\section{Methods \label{sec:methods}}

\begin{figure}
    \centering
    \includegraphics[width=0.32\linewidth]{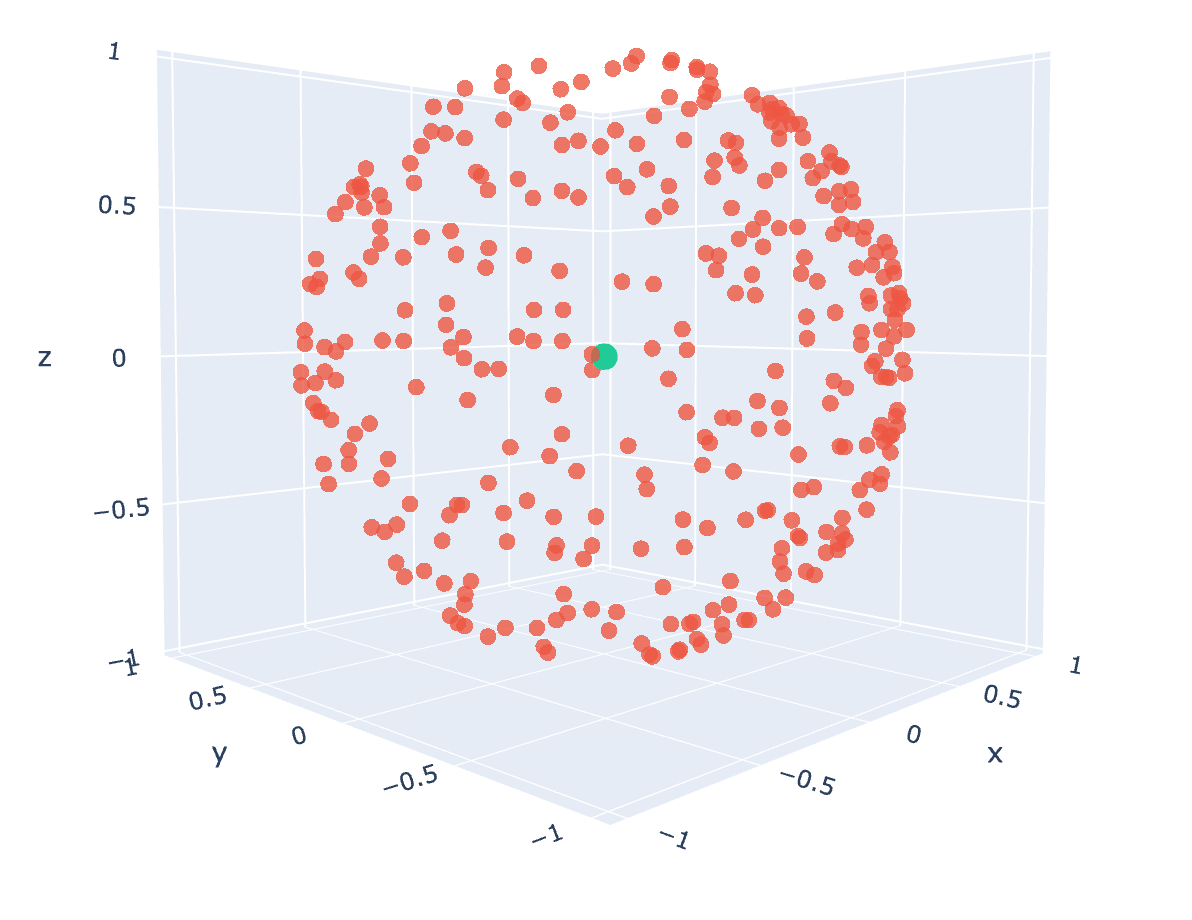}
    \includegraphics[width=0.65\linewidth]{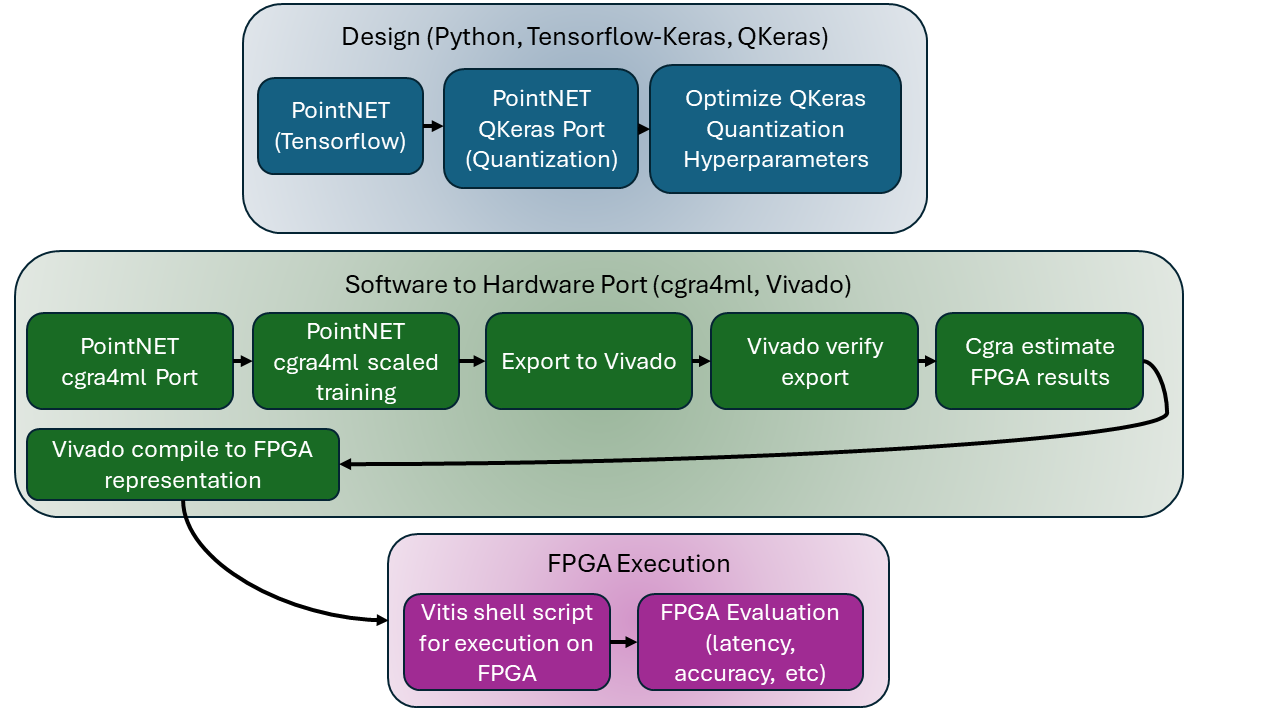}
    \caption{LEFT: A visual illustration of KamLAND-Zen's inner PMT Point Cloud Data. Image is originally from Reference \cite{fu_generate_2024}. RIGHT: Flow chart of FastPointNet. The top box shows the design and model testing flow. The middle box shows the flow from porting the PointNet model to the hardware. The bottom box shows the flow for executing and testing the hardware deployment of PointNet.}
    \label{fig:flowchart}
\end{figure}
An overview of the entire pipeline may be found in Figure \ref{fig:flowchart} RIGHT. The flow chart has three phases: 1) design phase, 2) software-to-hardware port phase, and 3) FPGA execution phase. Each phase includes various subprocedures as outlined in the flow chart. The first step was to design and train PointNet over a dataset of 100,000 KLZ simulated events \cite{klz_search_for_majorana_nature}. The shape of each point in the dataset is described in Section \ref{sec:intro}. In addition to the five dimensions, an additional binary trigger dimension is added based on whether or not the PMT received a light signal: when event energy is low, it is possible that some PMTs will not receive any light signal. To reconstruct the physics inside the detector, a mapping of these six features to four outputs is desired: the three spatial coordinates of the vertex of the physics event and the energy of the event. More rigorously, this model takes the form:
\begin{equation}
    f(\textbf{x}): \mathbb{R}^{6\times 2126}\rightarrow \mathbb{R}^4
    \label{eqn:func}
\end{equation}
The network architecture selected for this purpose was based on PointNet, an architecture particularly suited for data involving point clouds—data structures that preserve spatial semantics while remaining invariant to permutation \cite{qi_pointnet_2017}. Because the events in the dataset are point clouds, PointNet is a sound architecture for fitting Equation \ref{eqn:func}. 
%Following the flow in Figure \ref{fig:flowchart} RIGHT, the first step was to design and tune the PointNet model and verify that it fit the data well, which was accomplished using Tensorflow-Keras.
The PointNet model was built using Tensorflow-Keras \cite{tensorflow2015-whitepaper} with its training parameters listed in Table \ref{tab:design_phase_hyperparameters}. The implementation the PointNet model may be found in Appendix \ref{sec:tensorflow_model_design}.

The next step involved quantizing the PointNet model, which required defining the bit precision for the input, output, and layer parameters of the PointNet architecture. This step was critical due to the strict resource constraints of the FPGA, necessitating the compression of the model to minimize its physical utilization. Moreover, transitioning the machine learning model from software to hardware requires specifying the bit allocation for each layer of the FPGA. High-level quantization provides these parameters, enabling the fine-tuning of resource allocation for optimal performance without directly interfacing with the FPGA. A description of the implemented layers in the compressed model may be found in Appendix~\ref{sec:qkeras_model_implementation}. After training and quantizing the model using \texttt{QKeras} \cite{google_qkeras_nodate}, a hyperparameter search was conducted to obtain layer-level bit parameters. The hyperparameter search was executed to minimize the mean squared error (MSE) between the predicted and true event position and energy of each event in the validation split. The loss function used was an MSE function:
\begin{equation}
    \mathcal{L}=\mse(x)+\mse(y)+\mse(z)+\mse(\mathrm{energy})
    \label{eqn:loss_mse}
\end{equation}
The script for the hyperparameter search may be found in Appendix \ref{sec:hyperparam_search}.

% As mentioned earlier in section \ref{sec:methods}, quantizing the PointNet model yielded the number of bits required for each layer when deploying the model to the FPGA. Varying the number of bits in each layer directly contributes to the error associated with compressing the model, indicating the necessity of a hyperparameter search for optimal number of bits in each layer. The hyperparameter search is executed to minimize the mean squared error (MSE) between the predicted and true event position and energy of each event. The loss MSE function is:
% \begin{equation}
%     \mathcal{L}=\mse(x)+\mse(y)+\mse(z)+\mse(\mathrm{energy})
%     \label{eqn:loss_mse}
% \end{equation}

% \begin{enumerate}
%     \item Because the FPGA is strictly limited in the space it has for representing the model, the model needed to be compressed as much as possible to minimize the physical utilization of the FPGA.
%     \item When going from the software to hardware representations of the machine learning model, the FPGA needs to know how many bits to allocate for each layer. Quantizing the model at a high-level yields such parameters and allows us to tune these allocations for optimal performance without involving the FPGA directly.
% \end{enumerate}

Having established the quantized model and its optimal parameters, the next step was to port the model to a format that was compatible with the \texttt{cgra4ml} library, a software framework that facilitates the export of a complicated machine learning model to the AMD Vivado platform. Vivado then synthesized the model into an FPGA representation compatible with the \rfsoc FPGA development board. A similar \rfsoc{chip} will be adopted by the KamLAND-Zen experiment for their future upgrades. After synthesizing the model using the resources found in Table \ref{tab:fpga_deployment_resources}, AMD Vitis deployed the model onto the RFSoC 4$\times$2 board. The model's performance and reconstruction speed were estimated after the deployment step, and these results are described in Section~\ref{sec:results}. In summary, the pipeline involves three phases: the design and quantization phase, software to hardware port using \texttt{cgra4ml} and Vivado, and the execution and evaluation phase using Vitis.

\section{Results \label{sec:results}}
This section describes the three key results of deploying the PointNet model onto an FPGA. As discussed in Section~\ref{sec:methods}, quantizing the PointNet model yielded the number of bits required for each layer when deploying the model to the FPGA. The results from the hyperparameter search are shown in Table \ref{tab:quantization}. Using the results of this hyperparameter search, the optimal parameters for compression were selected: eight bits of integer precision and twelve bits of mantissa precision. The specific information regarding each layer may be found in Appendix \ref{sec:cgra_model_implementation}.

\begin{figure}
\begin{centering}
    \includegraphics[width=0.62\linewidth,trim={5pc 6pc 5pc 0pc},clip]{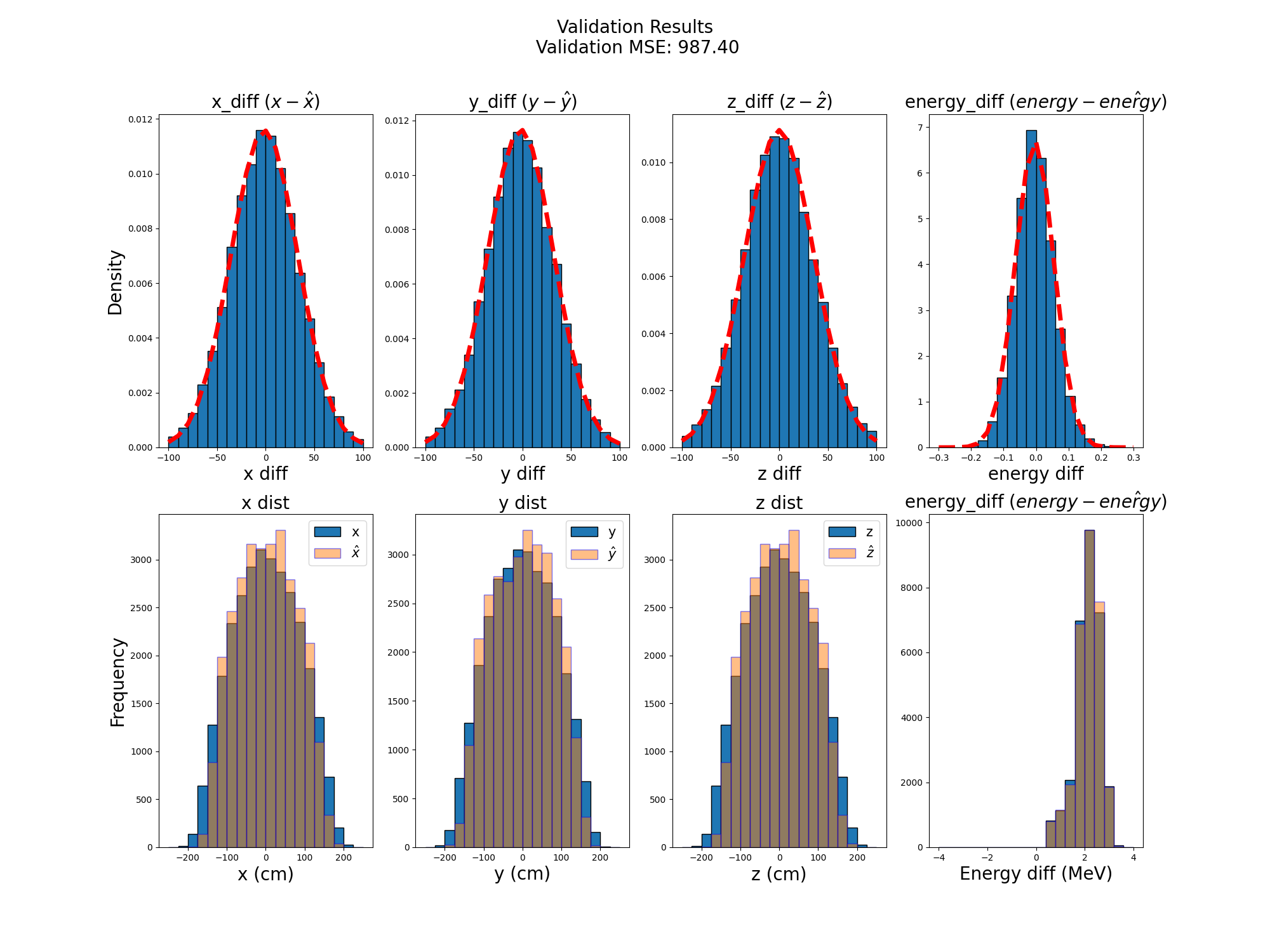}
    \includegraphics[width=0.34\linewidth,trim={0pc 0pc 0pc 0pc},clip]{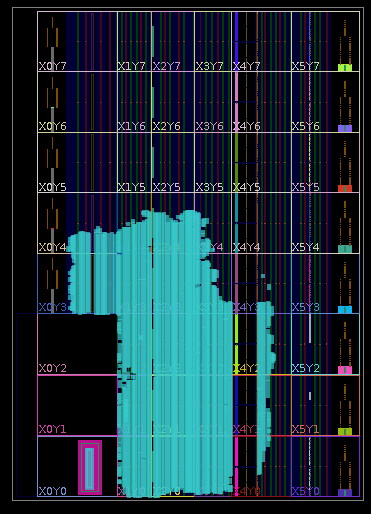}
    \end{centering}
    \caption{LEFT: Performance of \texttt{cgra4ml}-ported PointNet model. The bottom four panes show the reconstruction of the event's location and energy; the blue histogram shows the true position/energy from the simulated dataset while the yellow histogram shows the machine-learning-reconstructed positions and energy. The top four panes show the histograms of the difference in the prediction versus the true labels. The green lines show the Gaussian fit from the \texttt{SciPy} library. RIGHT: The Vivado synthesis of PointNet on the \rfsoc{FPGA}. The cyan-colored region shows the utilized FPGA resources on the chip.}
    \label{fig:accuracy}
\end{figure}

Having established the best kernel and bias quantization parameters, the next step was to port the PointNet model from Tensorflow-Keras to \texttt{cgr4ml}. Using \texttt{cgra4ml} and Vivado, the model was exported and synthesized for use in the \rfsoc board. The synthesized model is shown in Figure \ref{fig:accuracy} RIGHT. Training the ported model yielded the accuracy shown in Figure \ref{fig:accuracy} LEFT. Unfortunately, the optimal parameters found in the hyperparameter search were not used to train the ported model since they were incompatible with \texttt{cgra4ml}. The authors are working directly with the \texttt{cgra4ml} developers to deploy the most optimal model onto the FPGA as part of their future work. Instead, eight bits of mantissa precision were used with zero bits of integer precision.

From Figure \ref{fig:accuracy}, it is clear that the neural network was able to recover the spatial coordinates and energy of the event in distribution. The average MSE error across all batches during validation was 978.40. This error was higher than the results from the quantization step (366.25) since the best-performing model from quantization was incompatible with \texttt{cgra4ml}.
%; however, the \texttt{cgra4ml}-ported version exposes more parameters than the \texttt{QKeras} model that may be tuned for better performance. 
Table \ref{tab:all_training_results} summarizes the error for each label across different experiments, along with the nominal reconstruction error from the traditional method for comparison. A more detailed description of these errors may be found in Appendix~\ref{sec:model_prediction_error}. The \texttt{QKeras} model produced a spatial resolution close to the nominal values while outperforming the nominal energy resolution. The \texttt{cgra4ml}-ported model performed worse on position reconstruction yet outperformed the traditional method on energy reconstruction.

% Using the Gaussian fits from Figure \ref{fig:accuracy}, the error in x, y, and z is 35.6, 35.5, and 37.9, cm respectively. The calibration data from KLZ suggests a nominal radial resolution of $\pm 17\sqrt{3}$ cm \cite{Li:2020xxb}, meaning the current deployed resolution of $r\pm 62.96$ cm is worse than nominal\footnote{A more-detailed description of these errors may be found in Section \ref{sec:model_prediction_error}}.

\begin{table}[hbt!]
  \caption{Training Results}
  \label{tab:all_training_results}
  \centering
  \begin{tabular}{p{4cm}p{3cm}p{1cm}p{1cm}p{1cm}p{1cm}}
    \toprule
    % \multicolumn{2}{c}{Part}                   \\
    % \cmidrule(r){1-4}
    Experiment & Avg. Validation MSE & x Error (cm) & y Error (cm) & z Error (cm) & E Error (MeV) \\
    \midrule
    Traditional Method \cite{Li:2020xxb} & N/A & 17 & 17 & 17 & 0.14\\
    \texttt{QKeras} & 366.25 &  20 & 21 & 21 & 0.06\\
    \texttt{cgra4ml} & 987.40 & 34 & 34 & 36 & 0.06\\
    \bottomrule
  \end{tabular}
\end{table}

\subsection{FPGA Latency Results \label{sec:latency_results}}
The primary motivation for deploying PointNet onto an FPGA was to enable fast reconstruction; therefore, understanding the latency--defined as the time taken to obtain a result from the network after input—is crucial for evaluating the model's performance. The latency for the untrained model was 6,996.7 ms per batch and 6,980.9 ms per batch for the trained model. These values are averaged over 20 runs with a batch size of 16 events. This indicates a 436.3 ms inference speed per event. This represents a significant reduction in the time required to obtain key physics information, from approximately $\mathcal{O}(1~day)$ in offline reconstruction to $\mathcal{O}(1~s)$ in on-FPGA reconstruction. Future work will focus on optimizing the model to accelerate reconstruction speed to $\mathcal{O}(1~ms)$. Given that KamLAND-Zen collects roughly one data point per millisecond, achieving $\mathcal{O}(1~ms)$ reconstruction would allow for real-time event position and energy reconstruction as KLZ collects data. Considering this demonstration was performed on a single FPGA and that the next-generation KamLAND-Zen detector will consist of 120 FPGAs, achieving an inference time on the order of $\mathcal{O}(1~ms)$ is feasible with proper optimization.

%It is unsurprising that the latency between the trained and untrained model is approximately the same, as the FPGA's performance is dictated by the size of the model and not necessarily the individual layer weight values.

% \begin{figure}
%     \centering
%     \includegraphics[width=0.5\linewidth]{implemented design.png}
%     \caption{Utilization map of RF-SoC4x2 after Vivado synthesis. The light blue region is where the deployed model lives}
%     \label{fig:implemented_design}
% \end{figure}

\section{Summary \label{sec:summary}}
This paper presents a pipeline for deploying a PointNet model onto an FPGA for real-time event reconstruction of position and energy in the KamLAND-Zen experiment. While position reconstruction accuracy was worse compared to the traditional method, the energy resolution was improved. A key advantage of this approach is the significant reduction in latency to obtain reconstruction results from several days to just a few seconds. Future work will involve collaborating with the \texttt{cgra4ml} developers to port the best-performing model and optimizing the model structure for faster inference speed. In the long term, this framework will be integral to the deployment of 120 data acquisition boards in the next generation of the KamLAND-Zen experiment.
\section{Broader Impacts \label{sec:broader_impacts}}
In this section, the broader impacts of this work are discussed.
\subsection{Limitations \label{sec:limitations}}
This work has some limitations. One major limitation is that the model's accuracy was not tested against permutations of the dataset. Moreover, this model was only trained on one dataset. Therefore, to make the model more robust, a k-fold analysis of the accuracy would need to be completed. Another major limitation is that the model was trained on simulated detector physics, which may not accurately reflect the true reconstruction physics in the detector. While the deployed model's computational efficiency is unaffected by dataset size, it is affected by the number of inputs; increasing the number of inputs (PMTs) increases the number of compute resources necessary during both the design and deployment phases. This work is limited in its application and thus cannot be used for harmful acts.
\subsection{Societal Impacts}
The application of this work is limited to the context of the KLZ experiment, thus it cannot pose any societal harm. However, the models that are developed throughout the work are computationally expensive to train. Using shared compute resources that are energy-efficient may help mitigate the energy used to train these models. See Table \ref{tab:design_compute_resources} for the compute resources used in this work.

\begin{ack}
The KamLAND-Zen experiment is supported by JSPS KAKENHI Grants No. 21000001, No. 26104002, and No. 19H05803; the U.S. National Science Foundation awards no. 2110720 and no. 2012964; the Heising-Simons Foundation; the Dutch Research Council (NWO); and under the U.S. Department of Energy (DOE) Grant No. DE-AC02-05CH11231, as well as other DOE and NSF grants to individual institutions. The Kamioka Mining and Smelting Company has provided service for activities in the mine. We acknowledge the support of NII for SINET4. We also acknowledge support from the A3D3 Institute under grant number 2117997. The computational resources that enabled this work was provided by the San Diego Supercomputer Center Expanse cluster.  We thank  Javier Duarte, Abarajithan Gnaneswaran, Ryan Kastner, Elham Khoda, and Zhenghua Ma (alphabetically ordered) for the useful discussion and technical support on \texttt{cgra4ml}.
\end{ack}

\bibliography{neurips_2024}

\newpage
\appendix

\section{Appendix / supplemental material}

% \subsection{Propagated Error for Radial Resolution}
% In this section we give a brief derivation of the propagated error for the radial resolution. In our approach, we train the x, y, and z dimension separately. However, the spatial error from calibration is approximately symmetric in the KLZ experiment. To compare our error results with the KLZ calibration data, we must propagate the error on x, y, and z. The magnitude of the radius from x, y, and z is:
% \begin{equation}
%     r=f(x,y,z)=(x^2+y^2+z^2)^{1/2}
% \end{equation}
% The error on $r$ is found from error propagation:
% \begin{equation}
%     \delta{f}^2=(\delta{x})^2(\partial_x{f})^2+(\delta{y})^2(\partial_y{f})^2+(\delta{z})^2(\partial_z{f})^2
% \end{equation}
% This gives us the propagated error as:
% \begin{equation}
%     \delta r(x,y,z)\approx\sqrt{\frac{{\delta x}^2 x^2+{\delta y}^2 y^2+{\delta z}^2 z^2}{x^2+y^2+z^2}}
%     \label{eqn:r_prop_error}
% \end{equation}
% From pure geometry, the error on the radius is a function of x, y, and z. Therefore, to give an estimate of the radial error, we can use the mean error of each 

\subsection{Optimization Hyperparameter Search Script \label{sec:hyperparam_search}}
This is the script used to search for the optimal quantization parameters:
\begin{myverbatim}
for enc_r_bits in 8 12
do
    for enc_l_bits in 0 8 12
    do
        for o_bits in 0 8
        do
            python train_keras.py --epochs 50 \
            --save_ver "enc_r$enc_r_bits enc_l$enc_l_bits dec_r$enc_r_bits dec_l$enc_l_bits o_bits$o_bits" \
            --enc_a $enc_r_bits --enc_b $enc_l_bits --dec_a $enc_r_bits --dec_b $enc_l_bits --o_int_bits $o_bits
        done
    done
done
\end{myverbatim}

\subsection{Tensorflow Model Implementation \label{sec:tensorflow_model_design}}
Below is a layer-level description of the Sequential Tensorflow model implementation used to fit Equation \ref{eqn:func}:
\begin{enumerate}
    \item 1D Convolutional Layer (64 filters, kernel size 1)
    \item 10\% dropout layer
    \item ReLU layer
    \item 1D Convolutional Layer (64 filters, kernel size 1)
    \item 10\% dropout layer
    \item ReLU layer
    \item 1D Convolutional Layer (512 filters, kernel size 1)
    \item Global Average Pooling 1D
    \item Dense Layer (256 points, leaky ReLU activation)
    \item 0\% or 10\% Dropout Layer
    \item Dense Layer (64 points, leaky ReLU activation)
    \item Dense Layer (4 points)
\end{enumerate}

\subsection{QKeras Model Implementation \label{sec:qkeras_model_implementation}}
Below is a layer-level description of the Sequential Tensorflow \texttt{QKeras} model design:
\begin{enumerate}
    \item 1D Convolutional Layer (64 filters, kernel size 1, kernel and bias quantizers use \verb|quantized_relu|)
    \item Quantized Activation Layer (8 bits mantissa precision)
    \item 1D Convolutional Layer (64 filters, kernel size 1, kernel and bias quantizers use \verb|quantized_relu|)
    \item Quantized Activation Layer (8 bits mantissa precision)
    \item 1D Convolutional Layer (512 filters, kernel size 1, kernel and bias quantizers use \verb|quantized_relu|)
    \item Global Average Pooling 1D
    \item Dense Layer (256 points, leaky ReLU activation, kernel and bias quantizers use \verb|quantized_relu|)
    \item Quantized Activation Layer (8 bits mantissa precision)
    \item Dense Layer (64 points, leaky ReLU activation, kernel and bias quantizers use \verb|quantized_relu|)
    \item Quantized Activation Layer (8 bits mantissa precision)
    \item Dense Layer (4 points, kernel and bias quantizers use \verb|quantized_relu|)
\end{enumerate}

\subsection{\texttt{cgra4ml}-Ported Model Implementation \label{sec:cgra_model_implementation}}
Below is a description of the \texttt{XModel} implementation used in the \texttt{cgra4ml}-ported model. Note that the mantissa bit precision used across all layers is 8 for the input, 8 for the kernel, and 16 for the bias.
\begin{enumerate}
    \item XBundle. Core is XConvBN (0 integer bit precision for kernel and bias, 64 filters, 1 kernel size, activation is ReLU XActivation)
    \item XBundle. Core is XConvBN (0 integer bit precision for kernel and bias, 64 filters, 1 kernel size, activation is ReLU XActivation)
    \item XBundle. Core is XConvBN (0 integer bit precision for kernel and bias, 512 filters, 1 kernel size, activation is ReLU XActivation). Pool is Global Averaging with no activation. Flatten is true
    \item XBundle. Core is XDense (0 integer bit precision for kernel and bias, 256 points, activation is ReLU activation with negative slope of 0.125)
    \item XBundle. Core is XDense (0 integer bit precision for kernel and bias, 64 points, activation is ReLU activation with negative slope of 0.125)
    \item XBundle. Core is XDense (0 integer bit precision for kernel and bias, 6 points, no activation)
\end{enumerate}

\subsection{Model Prediction Error \label{sec:model_prediction_error}}
From the KLZ collaboration, the approximate nominal spatial resolution is reported to be approximately $\pm17$ cm in the z direction \cite{Li:2020xxb}, and this work assumes that this error is the same for the x and y dimensions as well. When reporting a model's error, the histogram of differences between the model's predicted and true labels of the x, y, z, and energy is considered: an example of this may be found in Fig. \ref{fig:accuracy} LEFT. Then, a Gaussian distribution is fitted onto this histogram using the \texttt{SciPy} Python package. From this fit, the 1-sigma result is reported for each spatial coordinate as well as the energy. In addition, the nominal energy resolution of KLZ is $0.72\%/\textrm{MeV}$ \cite{Li:2020xxb}. From Figure \ref{fig:accuracy} LEFT, the mean energy of the events from the dataset is approximately $2~\textrm{MeV}$; therefore, the KLZ 1-sigma energy resolution is approximated as $0.14~\textrm{MeV}$.

\subsection{Other Tables}

\begin{table}[hbtp]
  \caption{PointNet Model Card Based on Reference \cite{Mitchell_2019}}
  \label{tab:pointnet_model_card}
  \centering
  \begin{tabular}{@{}p{4cm}p{9cm}@{}}
    \toprule
    % \multicolumn{2}{c}{Part}                   \\
    % \cmidrule(r){1-4}
    Characteristic & Detail \\
    \midrule
    Authors & Charles r. Qi, Hao Su, Kaichun Mo, Leonidas J. Guibas \\
    Model Date & 2017 \\
    Training Algorithms & See appendices \ref{tab:design_phase_hyperparameters} and \ref{tab:cgra_port_phase_hyperparameters} \\
    Paper and Citation & See reference \cite{qi_pointnet_2017} \\
    Primary Intended Users & Fitting point cloud data \\
    Primary intended users & Researchers \\
    Evaluation factors & Total Mean-Squared-Error (MSE) between target and predicted x, y, z, and energy\\
    Dataset & Simulated dataset from the KLZ collaboration \\
    Data Motivation & Simulated data includes true labels of physics events, which is necessary for PointNet training \\
    Data Preprocessing & Inject data label "PMT label," which describes whether the PMT was triggered. PMT has label 0 or 1. Label 0 is registered if both charge and time labels are identically 0. Reshape data into (x, y, z, label, time, charge) format. \\
    Ethical Considerations & The way PointNet is applied is limited to the scope of the experiment and thus does not present ethical issues.\\
    \bottomrule
  \end{tabular}
\end{table}

\begin{table}[hbtp]
  \caption{Compute Resources Used for Design and \texttt{cgra} Port Phases}
  \label{tab:design_compute_resources}
  \centering
  \begin{tabular}{ll}
    \toprule
    % \multicolumn{2}{c}{Part}                   \\
    % \cmidrule(r){1-3}
    Hardware & Hardware Used \\
    \midrule
    Compute Name & UCSD Expanse Cluster \\
    GPU (Graphics Processing Unit) & NVIDIA Tesla V100 32GB VRAM \\
    CPU cores & 4\\
    RAM (Random Access Memory) & 64GB \\
    \bottomrule
  \end{tabular}
\end{table}

\begin{table}[hbtp]
  \caption{FPGA Deployment Computer Resources}
  \label{tab:fpga_deployment_resources}
  \centering
  \begin{tabular}{ll}
    \toprule
    % \multicolumn{2}{c}{Part}                   \\
    % \cmidrule(r){1-3}
    Hardware/Software & Hardware Used \\
    \midrule
    GPU (Graphics Processing Unit) & Radeon Graphics \\
    CPU & AMD Ryzen 7 5700G with Radeon Graphics\\
    RAM (Random Access Memory) & 12GB \\
    Storage & 500 GB \\
    Swap Memory & 8GB \\
    Vivado Version & 2022.2 \\
    Vitis Version & 2022.2 \\
    \bottomrule
  \end{tabular}
\end{table}

\begin{table}[hbtp]
  \caption{Quantization Phase Hyperparameters}
  \label{tab:design_phase_hyperparameters}
  \centering
  \begin{tabular}{@{}p{4cm}p{3cm}p{6cm}@{}}
    \toprule
    % \multicolumn{2}{c}{Part}                   \\
    % \cmidrule(r){1-3}
    Parameter & Value & Reasoning \\
    \midrule
    Epochs & 50 & Loss curve suggested that at least 30 epochs was necessary to reach convergence\\
    Batch Size & 128 & Larger batch size improved training results \\
    Learning rate & 1e-3 & Stable learning rate \\
    Training/Validation Split & 0.7/0.3 & N/A \\
    Optimizer & AdamW \cite{kingma2017adammethodstochasticoptimization} \cite{loshchilov2019decoupledweightdecayregularization} & Commonly-used optimizer \\
    \bottomrule
  \end{tabular}
\end{table}

\begin{table}[!hptb]
  \caption{\texttt{cgra4ml} Port Training Hyperparameters}
  \label{tab:cgra_port_phase_hyperparameters}
  \centering
  \begin{tabular}{@{}p{4cm}p{3cm}p{6cm}@{}}
    \toprule
    % \multicolumn{2}{c}{Part}                   \\
    % \cmidrule(r){1-3}
    Parameter & Value & Reasoning \\
    \midrule
    Epochs & 50 & Quantization phase also used 50 epochs \\
    Batch Size & 128 & Larger batch sizes would not train due to insufficient amount of RAM \\
    Learning rate & 1e-3 & Stable learning rate \\
    Training/Validation Split & 0.7/0.3 & ~ \\
    Optimizer & AdamW \cite{kingma2017adammethodstochasticoptimization} \cite{loshchilov2019decoupledweightdecayregularization} & Commonly-used optimizer \\
    \bottomrule
  \end{tabular}
\end{table}

\begin{table}[!hbtp]
  \caption{Implemented Assets and Licensing}
  \label{tab:implemented_assets}
  \centering
  \begin{tabular}{ll}
    \toprule
    % \multicolumn{2}{c}{Part}                   \\
    % \cmidrule(r){1-3}
    Asset & License or Permission \\
    \midrule
    AMD Vivado & University License \\
    AMD Vitis & University License \\
    \texttt{cgra4ml} & Apache-2.0 License \\
    KLZ Dataset & KLZ Collaboration \\
    \rfsoc Development Board & Board may be used for academic purposes only \\
    \bottomrule
  \end{tabular}
\end{table}

\begin{table}[!hbtp]
  \caption{Quantization Hyperparameter Results}
  \label{tab:quantization}
  \centering
  \begin{tabular}{@{}p{3cm}p{3cm}p{2cm}p{5cm}@{}}
    \toprule
    % \multicolumn{2}{c}{Part}                   \\
    % \cmidrule(r){1-3}
    Mantissa Precision & Integer Precision & Activation Integer Precision & Loss (MSE(x) + MSE(y) + MSE(z) + MSE(energy) \\
    \midrule
    8 & 0 & 0 & 404 \\
    8 & 0 & 8 & 9018 \\
    8 & 8 & 0 & 306 \\
    8 & 8 & 8 & 62099 \\
    8 & 12 & 0 & 368 \\
    8 & 12 & 8 & 18470 \\
    12 & 0 & 0 & 476 \\
    12 & 0 & 8 & 1670 \\
    12 & 8 & 0 & 362 \\
    12 & 8 & 8 & 19492 \\
    12 & 12 & 0 & 375 \\
    12 & 12 & 8 & 2265 \\
    \bottomrule
  \end{tabular}
\end{table}

\end{document}